\documentclass{emulateapj}
\usepackage{graphicx}
\usepackage{psfig}

\def\ltsima{$\; \buildrel < \over \sim \;$}
\def\simlt{\lower.5ex\hbox{\ltsima}}
\def\gtsima{$\; \buildrel > \over \sim \;$}
\def\simgt{\lower.5ex\hbox{\gtsima}}



\newcommand{\msun}{${\rm M_{\sun}}$}

\def\kms{{\rm\,km\,s^{-1}}}

\def\kpc{{\rm\,kpc}}

\def\msun{{\rm\,M_\odot}}

\def\AA{$\; \buildrel \circ \over {\rm A}$}

\slugcomment{{\sc Accepted in ApJ: }  }

\begin{document}

\title{A kinematically selected, metal-poor stellar halo in the outskirts of M31}
\author{S.\ C.\ Chapman\altaffilmark{1}, 
R.\ Ibata\altaffilmark{2}, 
G.\ F.\ Lewis\altaffilmark{3},   
A.\ M.\ N.\ Ferguson\altaffilmark{4}, 
M.\ Irwin\altaffilmark{5}, 
A.\ McConnachie\altaffilmark{6}, 
N.\ Tanvir\altaffilmark{7}
}
\altaffiltext{1}{California Institute of Technology, Pasadena, CA\,91125;
 \texttt{schapman@astro.caltech.edu}}
\altaffiltext{2}{
Observatoire de Strasbourg, 11, rue de l'Universit\'e, F-67000, Strasbourg,
France} 
\altaffiltext{3}{
Institute of Astronomy, School of Physics, A29, University of Sydney, NSW
2006, Australia}
\altaffiltext{4}{Institute for Astronomy, University of Edinburgh,
Edinburgh,EH19 3HJ, UK}
\altaffiltext{5}{
Institute of Astronomy, Madingley Road, Cambridge, CB3 0HA, U.K.}
\altaffiltext{6}{
University of Victoria, Dept. of Physics and Astronomy, 
Victoria, BC, V8P 1A1, Canada}
\altaffiltext{7}{
Physical Sciences, Univ. of Hertfordshire, Hatfield, AL10 9AB, UK}

\begin{abstract}
We present evidence for a metal-poor, [Fe/H]$\sim-1.4$\ $\sigma$=0.2 dex, stellar
halo component detectable at radii from 10\,kpc to 70\,kpc, 
in our nearest giant spiral neighbor, the Andromeda galaxy.
This metal-poor sample underlies the recently-discovered 
extended rotating component, and has no detected
metallicity gradient. This discovery uses a large sample of 9861
radial velocities of Red Giant Branch (RGB) stars obtained with the Keck-II
telescope and DEIMOS spectrograph, with 827 stars with robust radial velocity measurements
isolated kinematically to lie in
the halo component primarily by windowing out the extended rotating component
which dominates the photometric profile of Andromeda out to $<$50\,kpc
(de-projected).  The stars lie in 54 spectroscopic fields spread over an 8
square degree region, and are expected to fairly sample the halo
to a radius of $\sim$70\,kpc.  The halo sample shows no significant
evidence for rotation.  Fitting a simple model in which the velocity
dispersion of the component decreases with radius, we find a central
velocity dispersion of $152\kms$ decreasing by $-0.90\kms/\kpc$.  By fitting
a cosmologically-motivated NFW halo model to the halo stars we constrain
the virial mass of M31 to be greater than $9.0 \times 10^{11} \msun$ with
99\% confidence.  The properties of this halo component are very similar to
that found in our Milky Way, revealing that these roughly equal mass
galaxies may have led similar accretion and evolutionary paths in the early
Universe.
\end{abstract}


\section{Introduction}

One of the fundamental tenets of the Cold Dark Matter (CDM) paradigm
\citep{white78} is that dark matter halos form hierarchically, via a
series of mergers with smaller halos. This prediction gives rise to the
natural expectation that the stellar halo\footnote{We use {\it halo} to mean 
`related to dark halo',  and typically use {\it stellar halo} for the stellar component that partially 
fills the dark halo.}
is formed from disrupted, accreted
dwarf galaxies \citep{johnston99,helmi99,bullock01,bullock04}.

CDM models predict that stellar halos form early, with the majority of the
stars in the halo formed within a few relatively massive,
$5\times10^{10}$M$_\odot$, dwarf irregular (dIrr) -size dark matter halos.
These were accreted $\sim$10 Gyr in the past, likely prior 
to the formation of the
bulge component via gaseous collapse, mergers and starbursts.  Such models
predict that the halo should be metal-poor and relatively smooth within the
inner 50 kpc, where the tidal signatures of early stellar streams have been
erased \citep{bullock05}. The giant stellar streams in the Milky Way
\citep{ibata01a,newberg02} and Andromeda \citep{ibata01b, mcconnachie04}
testify to the longevity of minor accretions which occurred since the metal
poor halo formed.  The average stellar halo density profile is expected to
fall off with radius more quickly than that of the dark matter because the
stellar halo is formed from the most tightly bound material in infalling
systems, while the majority of the accreted dark matter is stripped and
deposited at larger radii.  The discovery of fossil evidence of these
accretion events via the identification of substructure within nearby
stellar halos may provide the only direct evidence that structure formation
is hierarchical on small-scales \citep{bullock05}.

One of the most promising systems for the detailed decomposition and 
study of the stellar halo and its relation to the dark matter halo
is the Andromeda galaxy, or M31, \citep{walterbos88, pritchet88,irwin05,geehan06,gilbert06}.  
Often considered the Milky Way's twin,
this galaxy, at a distance of $\sim$785~kpc, \citep{mcconnachie05}, lies
sufficiently close that modern instrumentation can be used to measure
spectra of individual stars in the top few magnitudes of the red giant branch (RGB), providing
access to their radial velocities and metallicities.  In this paper we probe
the outer stellar component of M31 using radial velocities and calcium II
triplet (CaT) metallicities of RGB stars to study the overall
kinematic characteristics.

\section{Observations and sample definition}

Since 2002 we have undertaken a spectroscopic survey of RGB stars (selected
by their optical colors and magnitudes) in M31 with the 10m W.\ M.\ Keck~II
telescope and the DEIMOS spectrograph \citep{faber03}.  The 54 spectroscopic
fields shown in Figure~1 comprise 9861 stars with measured radial velocities
(shown in Figure~2),
and represent our ongoing program to provide a complete kinematic and
metallicity characterization of the disk and halo of M31, as well as the 
accreted fragments that now litter this environment.
Details of the spectroscopic observations and the selection of M31 RGB stars
from the color-magnitude diagram (CMD) have been presented in previous
papers (\citealt{ibata05} -- I05; \citealt{mcconnachie03}; 
\citealt{ferguson02}).
Briefly, we used the standard DEIMOS multi-slit mode for low density outer fields,
but adopted our own {\it mini-slitlets} approach targeting $>600$ 
stars per mask in higher density inner fields
using small (1.5\arcsec\ long) slits, with data reduction using our own custom
software pipeline (I05). Stars are selected for masks based on the position of the 
red giant branch (RGB) in the color-magnitude diagram, with a limit of
$I<22$ for useful spectra.
For most fields, except where a structure like And-IX \citep{chapman05} 
or the \citet{huxor05} fuzzy clusters were targeted,
stars were given equal priority from the RGB tip to
$I$=21.25, then lower weight for fainter objects.
The S/N degrades to about 5 at the limit of useful data.
A summary of all fields observed with DEIMOS, the radial velocity yields, and 
the breakdown of stars by kinematics into halo, disk, and Galactic windows, 
is presented in Table~1.
Representative individual spectra are shown in Figure~\ref{repspec}, as a function 
of both RGB color and $I$-mag.

\begin{figure}[t]
\centerline{\psfig{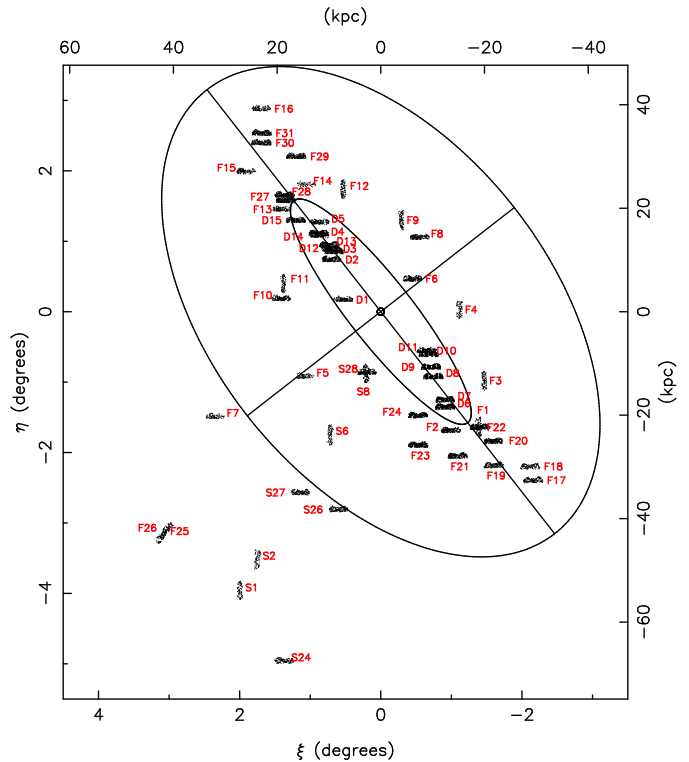}}
\vspace{6pt}
\caption{The current coverage of our radial velocity survey of M31 with the
DEIMOS spectrograph, in standard coordinates ($\chi, \eta$).  The outer
ellipse shows a segment of a $55\kpc$ radius ellipse flattened to $c/a =
0.6$, and the Major and minor axis are indicated with straight lines out to
this ellipse. The inner ellipse corresponds to a disk of radius $2^\circ$
($27\kpc$), with the same inclination as the galactic disk.  The
distribution of 54 Keck/DEIMOS fields, comprising 9861 stars with measured
radial velocities, are overlaid on the survey region, with field names tied to Table~1.  
}
\label{fig1}
\vspace{10pt}
\end{figure}

\begin{figure}[t]
\centerline{\psfig{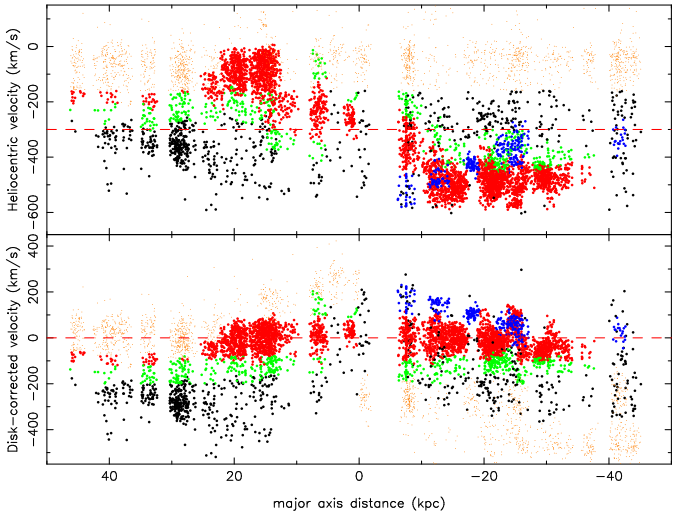}}
\vspace{6pt}
\caption{The radial velocity measurements from our 54 spectroscopic fields
are displayed as a function of distance along the Major axis, with the top
panel showing heliocentric values, and the lower panel velocities corrected
for the expected motion of the disk (using the model of I05). The foreground
Milky Way stars (shown with small orange dots) are removed from our sample.
The remaining stars are divided into a {\it disk-like} population
(disk-corrected velocity close to zero; red dots), an intermediate {\it
thick-disk-like} population (green dots), and a {\it halo-like} population
(a broad component which is not rotating with the disk; black dots), and
those metal-rich stars identified with the giant southern stream (blue
dots).  }
\label{vels}\vspace{10pt}
\end{figure}

\begin{figure}
\centerline{\psfig{figure=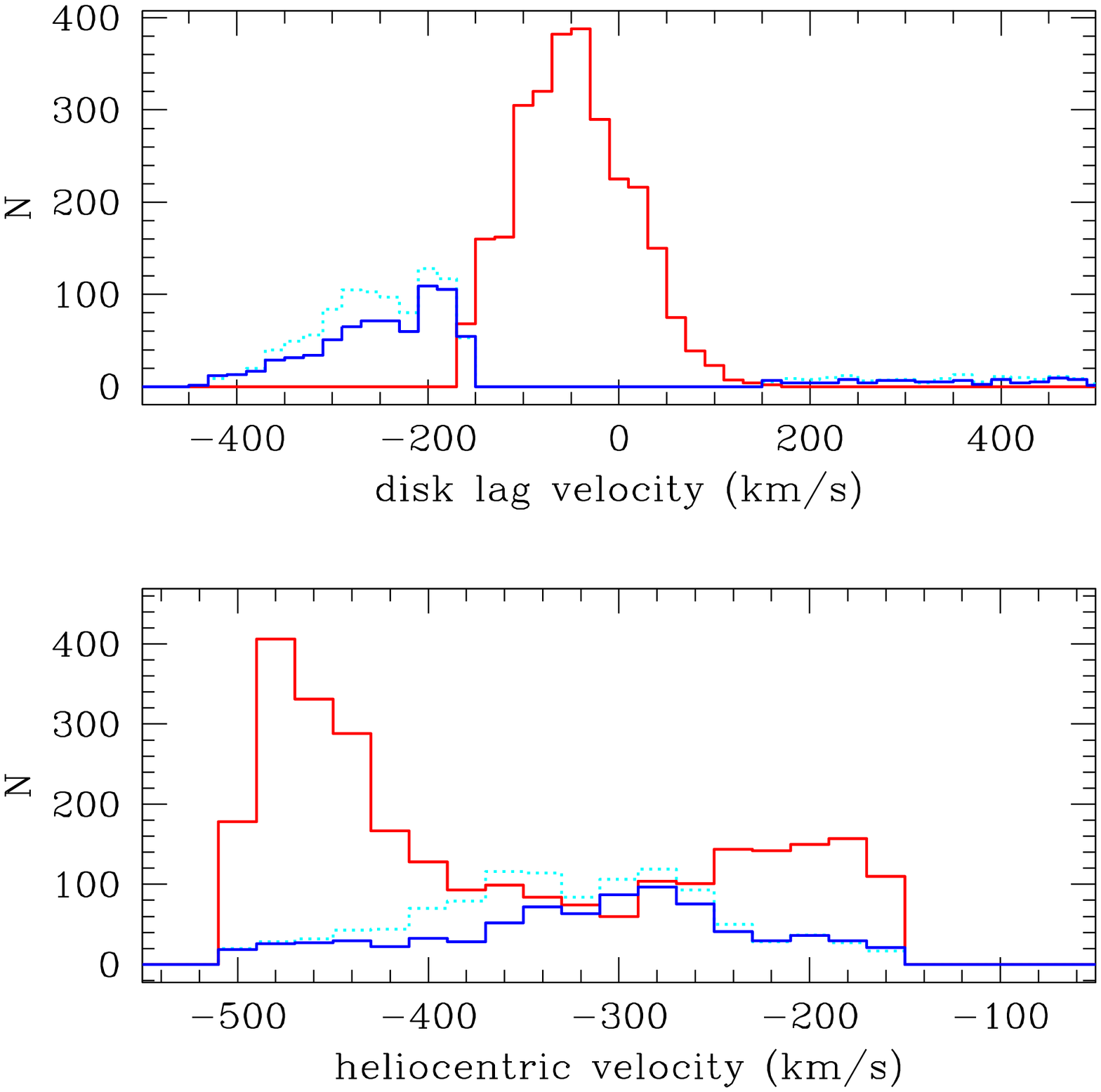,angle=0,width=3.5in}}
\centerline{\psfig{figure=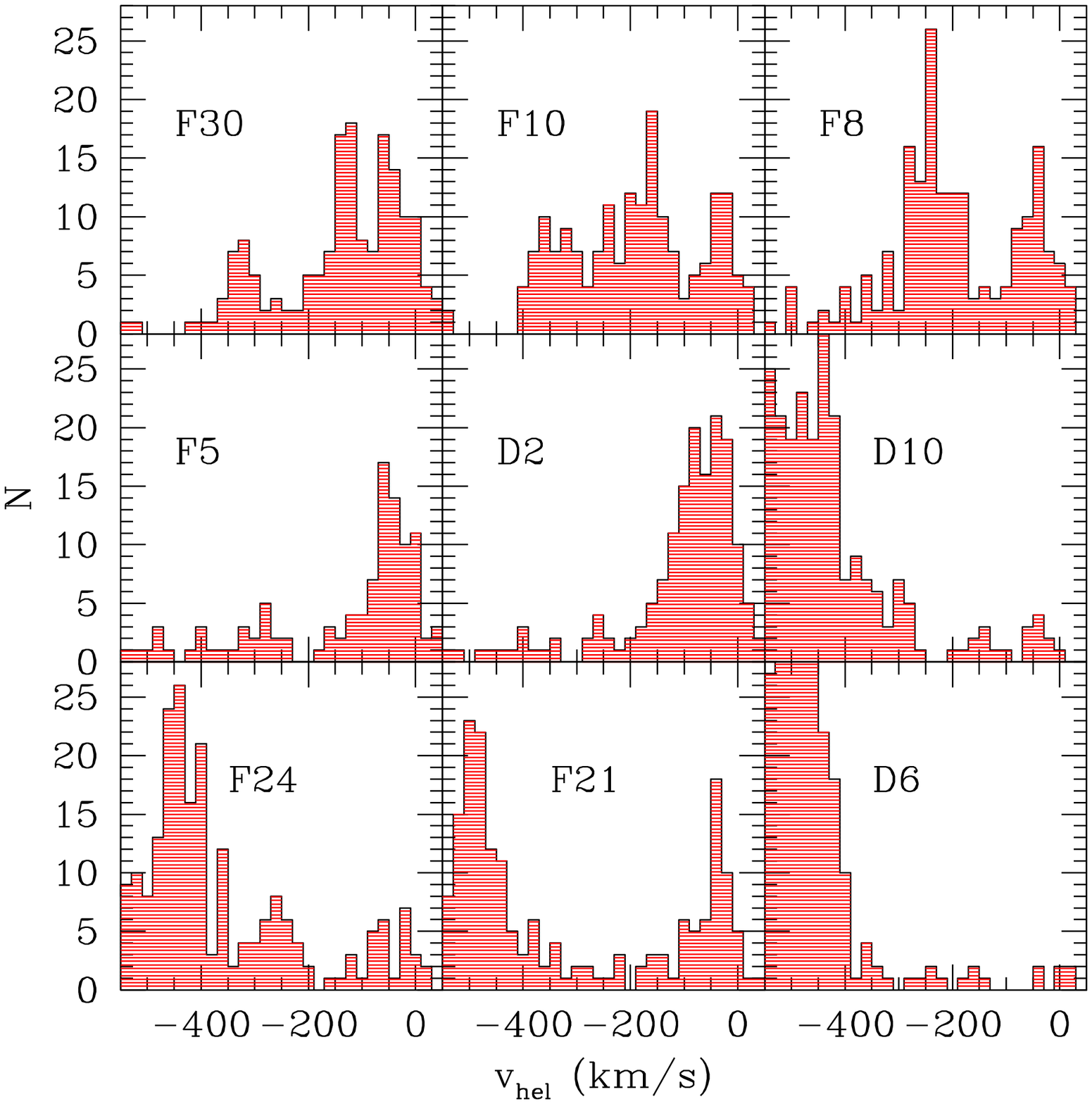,angle=0,width=3.5in}}
\vspace{6pt}
\caption{{\bf TOP:} Histograms of the halo/disk populations shown in Figure~\ref{vels}, where
{\it thin} and {\it thick} disk populations have been combined into a single 
histogram.  
{\bf Upper panel:} The kinematic sample selection, defined in the frame of {\it
disk-lag} velocities, with the {\it disk-like} population (shown in red)
centered at zero. 
The {\it halo-like} population
is chosen to be effectively non-rotating (v$_{disk-lag}>160$\,km/s).  The
halo sample is shown both with (dotted cyan) and without (solid blue) the
metal-rich kinematically coherent structure in the North-East region, which is
likely to be associated with the giant southern stream \citep{chapman06}.
{\bf Lower panel:} The same samples are shown in the heliocentric frame, highlighting the 
broad halo-like Gaussian distribution centered on the M31 systemic velocity.  
{\bf BOTTOM:}
Representative histograms of 9 individual fields spread throughout M31 (labeled as in 
Fig.~\ref{fig1}), and arranged roughly by location in M31. The double-peaked
disk/Galaxy components in the top three panels (from the N-E region of M31 -- 
F30, F10, F8) reveal that our v$_{hel}<-160$\,km/s Galactic windowing is indeed
very conservative, but deemed necessary to procure a nearly pure halo sample.
} 
\label{histos}
\vspace{10pt}
\end{figure}

\begin{figure*}[t]
\centerline{
\psfig{figure=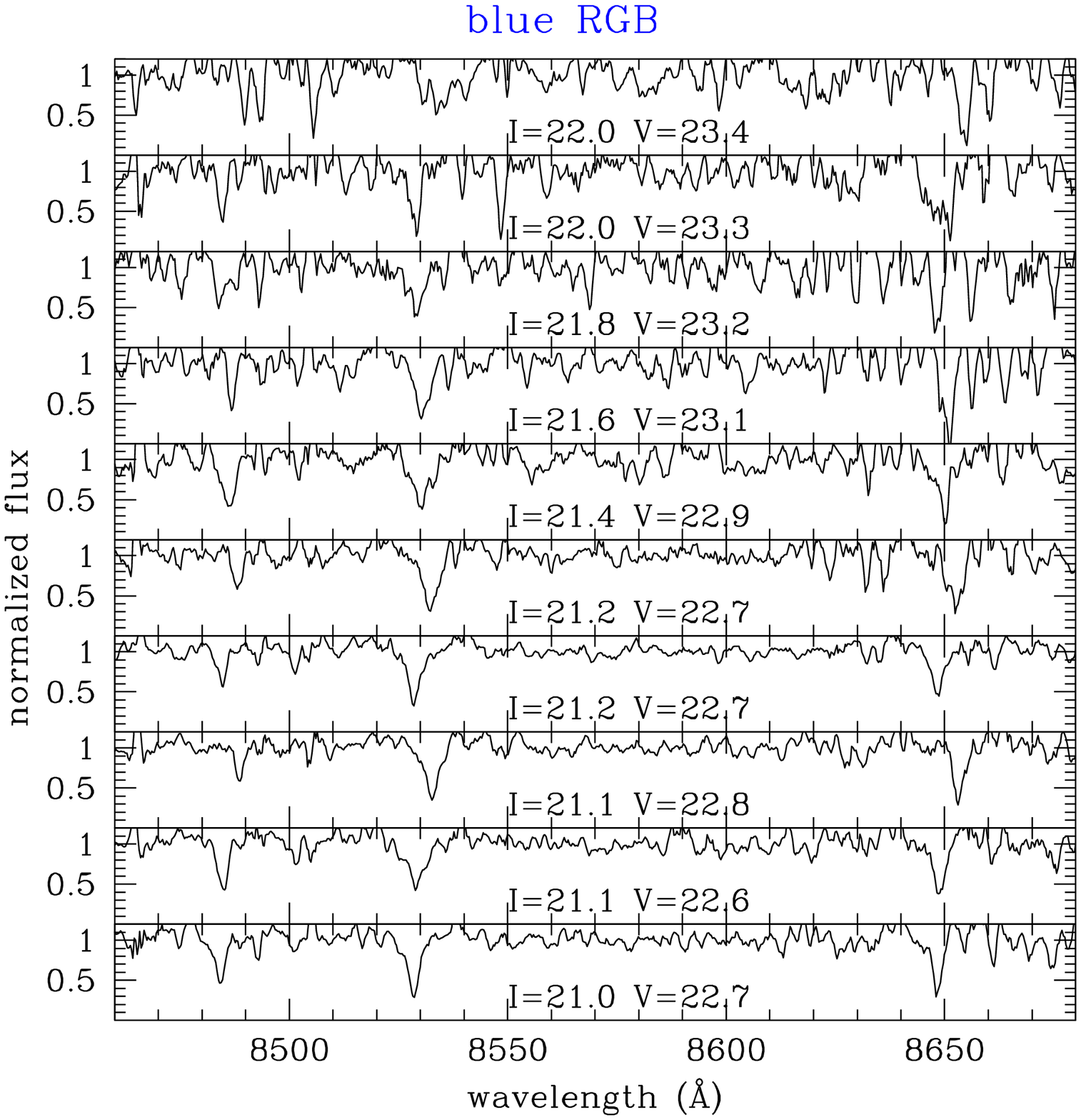,angle=0,width=3.5in}
\psfig{figure=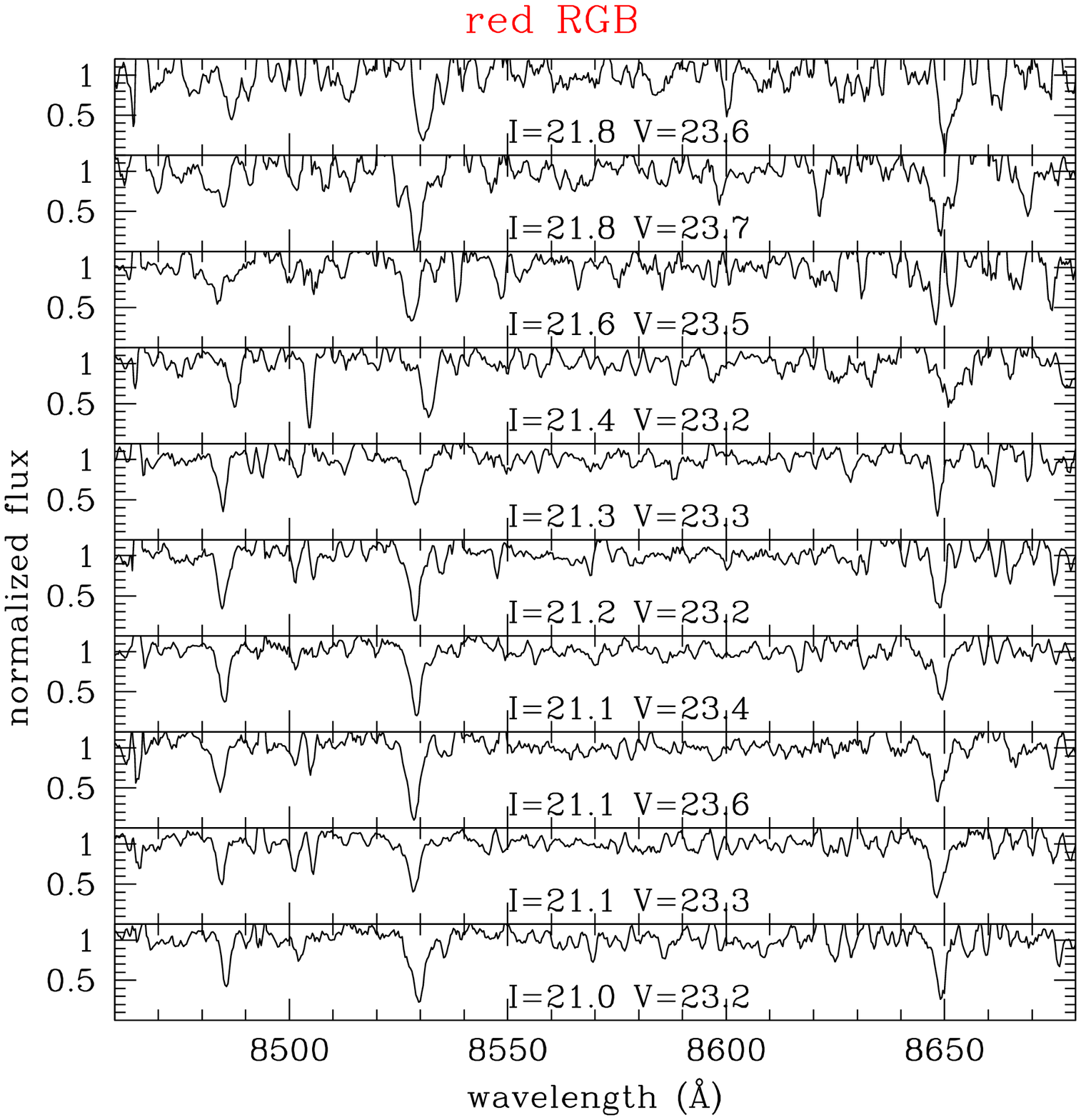,angle=0,width=3.5in}
}
\caption{
Representative spectra chosen from one of our survey fields (F3-w72), showing
blue RGB stars with $V-I=1.3$--$1.8$ (left) and red RGB stars with $V-I=1.8$--$2.5$ 
(right), over the range of magnitudes
for which we obtain good cross-correlations with the CaT template ($I<22$).
In this field there are no RGB candidates brighter than $I=21$, nor any good 
cross-correlations for stars fainter than $I=22$.
}
\label{repspec}
\vspace{10pt}
\end{figure*}

Radial velocities measured from the calcium II triplet (CaT) can be used to
kinematically isolate a likely sample of M31 halo stars.  
However, extracting a representative sample of
M31 halo stars is hampered by the fact that many of the spectroscopic
pointings have directly targeted the obvious photometric overdensities
identified in \citet{ferguson02}.  Fortunately, our discovery (I05) that an
extended rotating component dominates the star counts in all these
photometric substructures (with the exception of the giant tidal stream to
the South-East -- \citealt{ibata01b, mcconnachie03, ibata04, chapman06,raja06a,
kalirai06a}), 
suggests a method to
procure a sample of stars which is unbiased with respect to the photometric
overdensities.  A candidate {\it halo} component of stars can be
extracted by windowing out in heliocentric velocity all stars which rotate
with this extended component (shown in Figure~2).  In other words, the
photometric representation of our windowed sample should be spatially smooth
to first order (see I05 for further discussion of this point).  We construct
windows for the extended rotating component by adopting the average fit to
the extended rotating component, $\sigma_{v}=50$\,km/s Gaussian, from the
kinematic model presented in I05, and considering the window to span
$\pm2\sigma$ (or $\pm$100\,km/s in the {\it disk-lag} frame).\footnote{While 
the dispersion of the disk-like component is about 30\,km/s on average in 
individual fields (I05), it is broader ($\sim$50\,km/s) in the coadded disk fields.
This is partly a problem of  the projection effects and the
simplicity of the kinematic model currently adopted.} 
This is considered to be our ``disk'' sample, those stars exhibiting 
thin-disk-like kinematics. Between disk-lag velocities of $-160 <
v_{lag} < -100\kms$ we define an additional component of intermediate
kinematics, which we term ``thick disk'' in the subsequent discussion (to be discussed
in detail in a subsequent paper). As the
peak of the velocity distribution of the disk is observed to shift slightly
between fields (typically by $20$ to $30\kms$) compared to the simple disk
model of I05, the selection windows are adjusted accordingly by a small
amount (hence the limits of the ``disk'' and ``thick disk'' selections in
Figure~2 are slightly ragged).
The disk-like population is revealed in Figure~\ref{histos} as a Gaussian
with disk-lag velocity of zero. Some skew in the distribution 
arises from the truncation in the North-East region
of M31 imposed in order to reject contaminating stars from the Milky Way. This also
affects the v$_{hel}$ visualization of the disk in Figure~\ref{histos}, 
giving rise to a much larger -470\,km/s disk peak
(whereas the disk is actually sampled with roughly even numbers from our spectroscopy).

The minor axis fields require an additional windowing criterion.
For inner fields where the rotating components dominate, we have
windowed out the ``disk'' component as it passes over v$_{hel}$=-300km/s (even though
this removes any possible halo sample as well -- clearly this is
unavoidable).
For outer fields where the halo dominates, we do not window at all.
The first such field is F5 at 20kpc on the SE minor axis (Figure~1), 
and all fields beyond that were considered to be ``pure halo''.

We do not explicitly distinguish a bulge component in this study, as it does not obviously
distinguish itself in our predominantly major axis fields.
We note that a relatively metal-rich {\it bulge-like} 
component is reported at minor axis radii from 12-20\,kpc by \citet{kalirai06},
consistent with other photometric studies of RGB stars at these radii
(Durrell, Harris, \& Pritchet 1994, 2001; Mould \& Kristian 1986;
Brown et al.\ 2003; Bellazzini et al.\ 2003; Worthey et al.\ 2005).
However, this is tangential to the present study.
Very few of  our fields lie within this radial range, and even fewer are used for measuring 
the halo spectroscopic metallicity (see \S4 and Fig.~9 
which shows the fields where the numbers and quality of the spectra are sufficient 
for obtaining a halo metallicity measurement).
As discussed above, our first minor axis field suitable for direct comparison with the
\citet{kalirai06} bulge is F5 at 20\,kpc. 
This field has a halo sample metallicity of [Fe/H] = -1.18 (to be presented in Fig.9),
and is consistent with the findings of \citet{kalirai06} within errors of their 
bridging halo/bulge fields.
To ensure that this result is not biased by Galactic dwarf star contamination,
we have tested the field F5 with velocity cuts at both -300\,km/s and -170\,km/s, 
The resulting measurements are [Fe/H] = -1.23 ($v<-300$\,kms)  and [Fe/H] = -1.18 
($v<-170$\,kms).
This is in the opposite sense from that expected if dwarf contaminants were significant
(see below).
Our current inability to clearly distinguish a possible bulge component in  
the inner-most regions of M31 may bias our measurements to slightly higher metallicity,
although there is no statistically significant evidence that this is the case.

The Galactic foreground (discussed below) and the Giant Southern Stream are the last {\it 
contaminants} to window from our stellar halo sample.
The Giant Southern Stream has kinematics which
stand out as a high negative velocity peak in Figure~2 \citep{ibata04}.
It appears in only seven of the fields, although another
eight fields in the North-East region show a coherent kinematic structure
which is similarly metal-rich to the Giant Southern Stream, suggesting that
this region represents a dominant component of the Stream's {\it return}
on the other side of M31 \citep{chapman06}.
Nonetheless, the total number of stars in the Giant Stream is still relatively small, 
leaving the extended rotating
component windows having the most significant effect on the halo sample.

The sample selection in the disk-lag and heliocentric frames is shown in
Figure~\ref{vels} (where the black dots mark the halo stars of interest to this
study) and Figure~\ref{histos} where the total distributions are shown.
Figure~\ref{histos} reveals the stellar halo windowed sample as a 
broad Gaussian distribution centered on 
v$_{hel}\sim$-300km/s, along with the rotating thin and thick disk-like populations
together as a single sample.  
Also shown in Figure~\ref{histos} 
are representative velocity histograms in nine individual fields 
spread throughout M31. The double-peaked
disk/Galaxy components seen in the top three panels (from the N-E region of M31 -- 
F30, F10, F8) demonstrate that our v$_{hel}<-160$\,km/s Galactic windowing is indeed
very conservative, yet necessary to obtain an uncontaminated stellar halo sample.

After all windowing procedures, 
1207 stars remain in our halo sample. We then refine this sample to 827 stars
by applying the following quality criteria (the same criteria as in I05):
{\it (i)} The velocity of stars derived from two different sky-subtraction 
algorithms has to agree within 20km/s : removes 35 stars
{\it (ii)} Stars with continuum $> 10$ counts (or about 40\,e- total/pix) : removes 38 stars
{\it (iii)} Stars with $\sigma_v < 20$km/s : removes 138 stars
{\it (iv)} Stars with cross-correlation peak $< 0.05$ 
: removes 189 stars.
This sample can be considered to {\it underlie} the
extended rotating structure of (I05).

\subsection{Galactic dwarf contamination}

To complete our halo sampling, we must remove the foreground Galactic
component (mostly dwarf stars), observed as a strong Gaussian peak centered
on -61\, km/s (I05).  The Galactic
window is stationary (defined here as stars with $v_{hel} > -160 \kms$), except
for fields within the dense disk of M31 on the North-Eastern Major axis
where we chose to remove only those stars with $v_{hel}> 0 \kms$ as the
source density is so high that Galactic contaminants are rare.  
Any Galactic contamination near the M31 systemic velocity ($v_{hel} = -300 \kms$)
has to be Galactic halo, and the corresponding density of dwarf stars 
is minimal.
The Besan{\c c}con Galactic  starcounts model \citep{robin} 
estimates 10 Galactic stars per M31 DEIMOS field of view ($16.7\arcmin \times 
5\arcmin$) 
with  $v_{hel} < -160 \kms$) if all available candidates were observed.
For $v_{hel} < -300 \kms$ the Besan{\c c}on model indicates the number of
Galactic stars per DEIMOS field to be only 1.8 stars (typically $<<$1\%).
The Besan{\c c}on prediction at velocities $<$ -300 km/s is particularly relevant 
since more than 50\% of our halo sample is cut at v$_{hel}<$-300\,km/s.
In practice, we have typically observed $<$50\% of the stars which are 
consistent with M31 RGB colors 
available in the region covered by a DEIMOS mask.
Given our choice of thousands of targets per field inside the inner ellipse in
Figure~1 (and a maximum multiplexing of $\sim 600$ stars per DEIMOS mask
with our {\it mega-holes} approach -- see I05 for further discussion), 
we estimate a negligible Galactic contamination out to $\sim$30\,kpc, 
and at most 10\% contamination in our fields at larger radius.
We also refer to the work of \citet{gilbert06} which uses a multi-dimensional rejection
of Galactic dwarfs based on a variety of spectral features (most of which are unavailable
to us, given the small extracted wavelength regime from our custom data reduction).
\citet{gilbert06} are therefore able to calibrate the use of kinematic selection alone
as a means to reject Galactic dwarfs, finding that $<$1\% of stars cut at v$_{hel}<-300$km/s
will be dwarfs, in agreement with the Besan{\c c}on model. 
Therefore our CMD selection of RGB stars coupled with kinematic sorting of the
candidate M31 stars by radial velocity is likely to remove the vast majority
of Galactic contaminants.

The windowing of stellar halo sample provides another check on Galactic contamination.
Our North-Eastern stellar halo represents a more 
`Galactic dwarf free' sample as it
is cut in velocity below -300\,km/s, whereas the South-Western stellar halo
is cut typically from -160\,km/s$<v_{hel}<$-300\,km/s.
As we might expect based on the \citet{gilbert06} work, we find almost
no differences between our North-East and South-West hemisphere
results, neither in kinematics (\S 3) nor in metallicity (\S 4),
and thus it is clear that Galactic dwarfs are not significantly affecting     
our stellar halo results.

\begin{figure*}[t] 
\centerline{
\psfig{figure=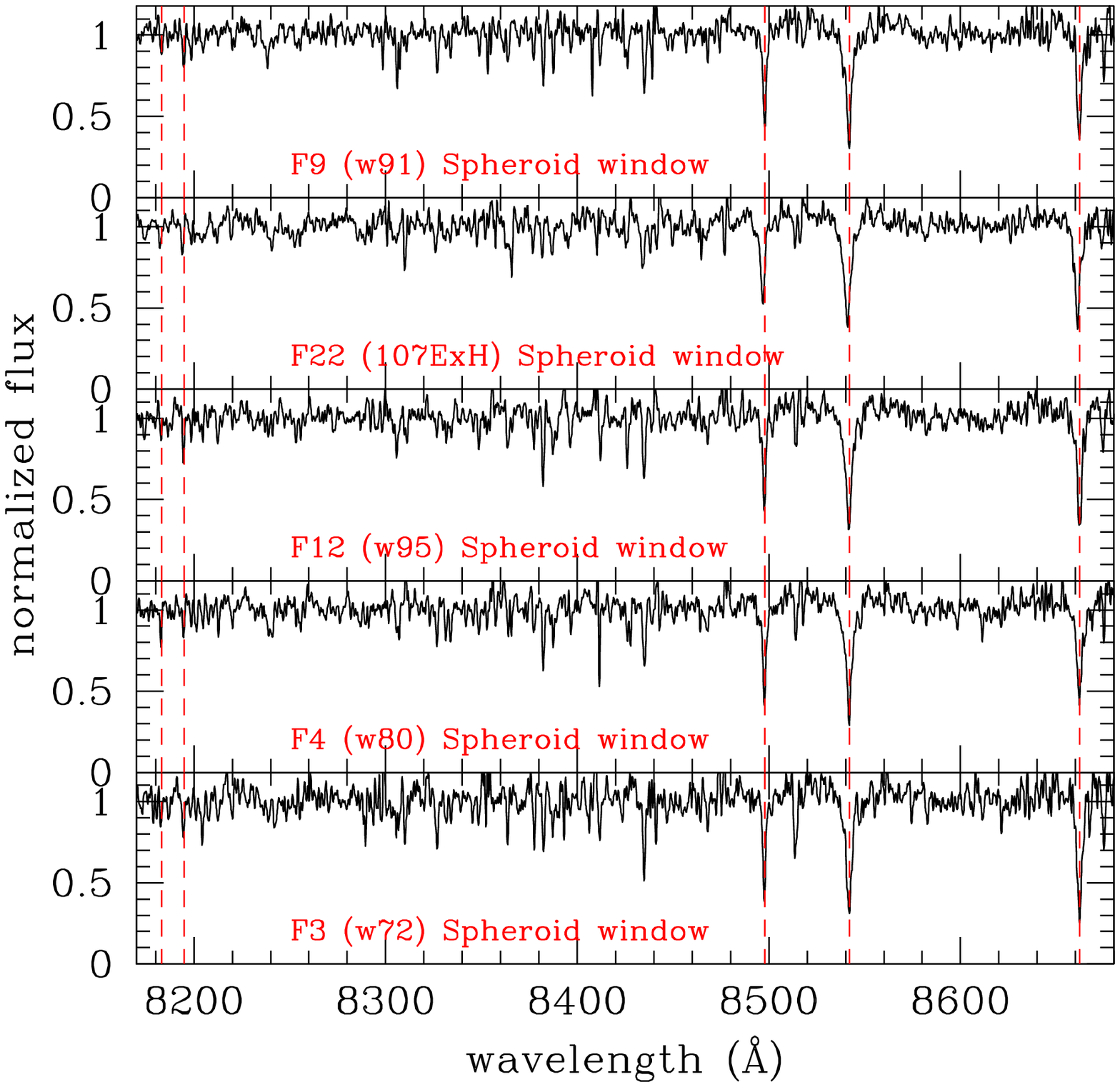,angle=0,width=3.5in}
\psfig{figure=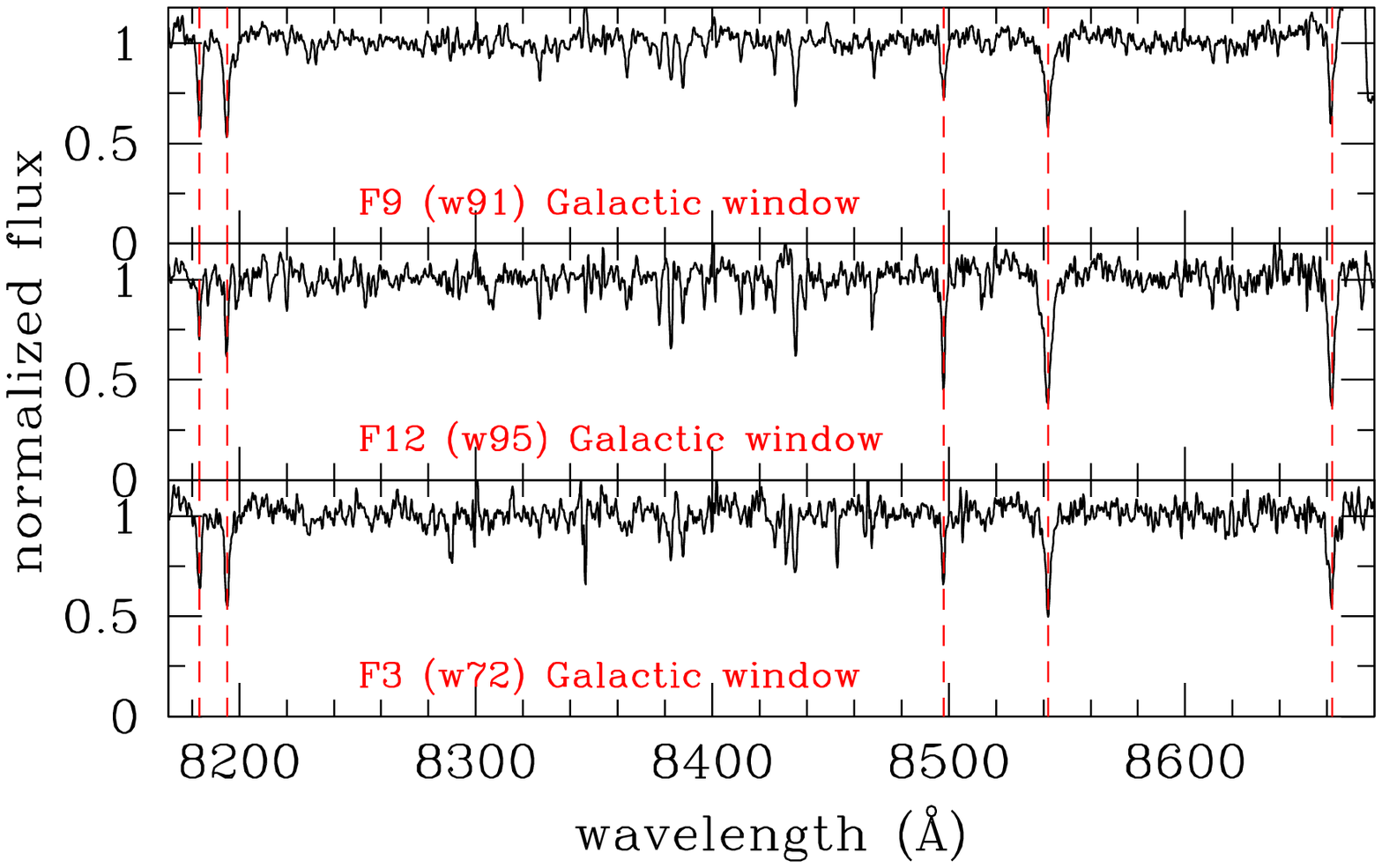,angle=0,width=3.5in}
}
\caption{Averaged spectra with wavelength coverage below 8200\AA\ 
for typical DEIMOS fields. Left panels show halo samples from both North-East and
South-West regions of M3, and therefore having velocity windows ranging from
$-160$\,km/s $>$ v$_{hel}>-300$\,km/s (F3,F4) to 
$-300$\,km/s $>$ v$_{hel}>-500$\,km/s (F12,F9,F22).
These fields show strong CaT absorption and minimal 
{Na\sc i}$_{\lambda8183,8195}$ absorption. Similar results are found
in all halo samples, providing a reality check on the minimal Galactic contamination to 
these samples at high negative velocities. 
The right panel shows the average of 
stars drawn from the v$_{hel}<-100\kms$ region thought to be highly contaminated
by Galactic dwarf stars in three of the same fields.
These stars show a strong {Na\sc i}
absorption typical of Galactic dwarfs with redder colors (e.g., Guhathakurta et al.\ 2006), 
with weaker CaT absorption than seen in the metal-poor halo samples.
There are between 10 and 16 stars in each stack.
}
\label{galcontam}
\vspace{10pt}
\end{figure*}

A further reality check that our spectroscopically identified stars are truly M31
RGB stars comes from the {Na\sc i}$_{\lambda8183,8195}$ which is 
sensitive to surface gravity, and is accordingly undetectable in M31
RGB stars, but is as strong as the CaT absorption lines in Galactic dwarfs 
with relatively red colors (e.g., \citet{raja06a}). 
We can check for this on a field by field basis in combined spectra 
of stars, and where our spectroscopic approach resulted
in wavelength coverage to $<8200$\AA\ 
(the $\sim$30\% of fields where we employed the standard DEIMOS multi-slit
masks). 
Our stacked spectra in all M31 components 
(extended disk, thick disk, halo, and Giant Stream) show very little evidence
for {Na\sc i}$_{\lambda8183,8195}$ absorption lines,
whereas stars drawn randomly from -100 to zero 
velocity show strong {Na\sc i} absorption on average (Figure~\ref{galcontam}).
Direct measurement of the {Na\sc i} equivalent widths in the halo and Galactic
windows shows the contamination fraction is less than 15\%, although this is of course a 
severe upper limit since there are large numbers of true M31 RGBs
in the ``Galactic'' window \citep{gilbert06} diluting the {Na\sc i} absorption, 
and weak {Na\sc i} is also observed in RGBs..
This further corroborates the very 
low Galactic contamination in our M31 kinematically-selected samples.

\section{Results: Kinematic properties of the stellar halo}

\begin{figure}[t]
\centerline{\psfig{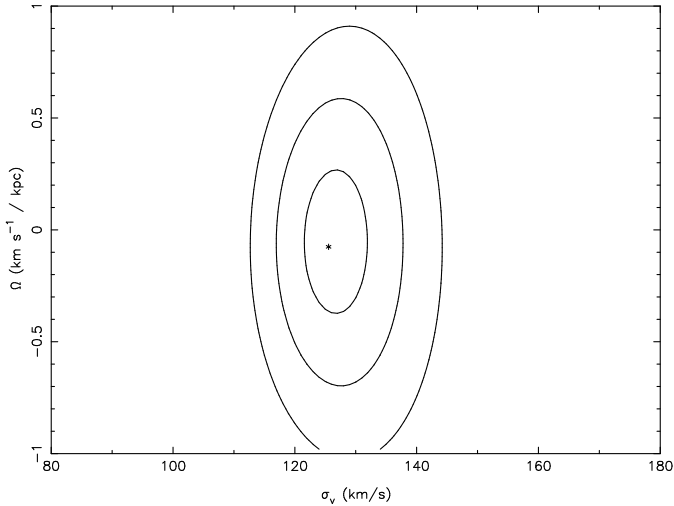}}
\vspace{6pt}
\caption{Likelihood contours (at $1\sigma$, $2\sigma$ and $3\sigma$
intervals) of a simple rotating isothermal halo model, as a function of
velocity dispersion and rotation. The asterix symbol denotes the position of
the most likely parameters $\sigma_v=126\kms$, $\Omega= -0.08 \kms /\kpc$,
where positive $\Omega$ denotes motion in the same sense as the galactic disk.
The data are clearly consistent with zero halo rotation at the $1\sigma$
level.}
\vspace{10pt}
\label{halofit1}
\end{figure}

To provide an overall kinematic characterization of the M31 halo stars, we first
fitted a simple model with an isotropic velocity dispersion that is
independent of radius, but that is allowed to rotate as a solid body about
the centre of the galaxy, which is assumed to have a systemic heliocentric
velocity of $-300 \kms$, and with the same axis of rotation as the disk.
For this first fit, we manually removed 3 fields from the sample, due to the
obvious presence of a kinematic halo substructure. These three contiguous
fields lie on the North-Eastern major axis at the edge of the inner ellipse shown
in Figure~1, and the kinematic substructure is clearly visible on Figures~2 \& 
\ref{histos} 
with heliocentric velocity in the range $-400 < v_h < -350 \kms$ (this detection is
is discussed further in \citet{chapman06}).
The model was fit simultaneously to the halo stars in all of the remaining 51 spectroscopic
field using a windowed maximum likelihood approach.  The resulting
distribution of likelihood as a function of the model dispersion and angular
speed is shown in Figure~\ref{halofit1}. The best model parameters given this data are a
halo velocity dispersion of $\sigma_v=126\kms$, and an angular velocity
of $\Omega= -0.08 \kms /\kpc$. However, the data are also consistent, at the
$1\sigma$ confidence level, with no rotation.

Indeed, the halo shows no significant signs of rotation in any
orientation.  By dividing the spectroscopic fields into two sub-samples
using a line intersecting the center of M31, we find the largest velocity
differential as a function of azimuthal angle occurs near the Major axis
where the NE half shows v$_{hel}= -292\pm8$\,km/s from a best fit Gaussian
to the average sample, while the SW half has v$_{hel}= -303\pm6$\,km/s.
This insignificant {\it rotation} is in the same sense as the extended
rotating component \citep{ibata05}, and could for instance be a result of
residual extended rotating component stars contaminating our halo
sample.  For simplicity we therefore neglect rotation in the following,
slightly more complex models.

\begin{figure}
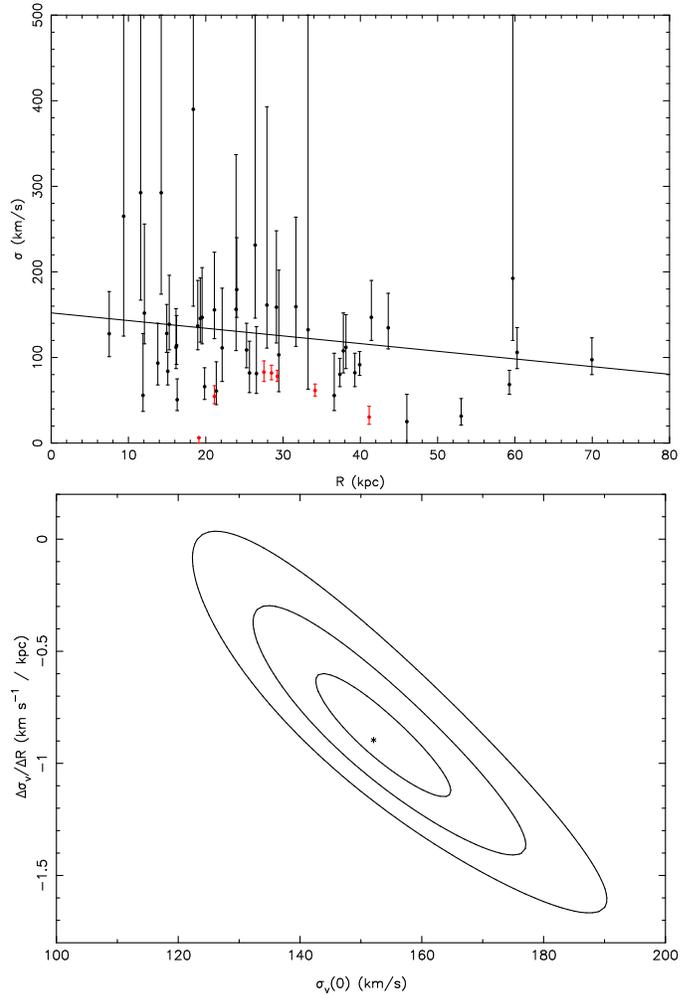

\vspace{6pt}
\centerline{\psfig{figure=f7a.ps,angle=-90,width=3.5in}}
\centerline{\psfig{figure=f7b.ps,angle=-90,width=3.5in}}
\vspace{6pt}
\caption{{\bf Upper panel:} Explicit single dispersion 
fits to the halo stars in the DEIMOS fields as a function of projected radius $R$. 
The vertical bars show $1\sigma$ uncertainties in these maximum-likelihood fits.
The substantial variations in the fitted values and their uncertainties
are to a large extent due to the low number of halo stars in most fields, and
due to the different window functions in each field.
{\bf Lower panel:} 
Likelihood contours (at $1\sigma$, $2\sigma$ and $3\sigma$
intervals) for a non-rotating halo model in which the velocity dispersion is
allowed to decrease linearly with projected radius $R$. The most likely
central velocity dispersion of this model is $\sigma_v=152\kms$ which
decreases with radius $R$ as $-0.90 \kms/\kpc$ (the straight line corresponding to these values is 
displayed in the upper panel). The fields in the upper panel that have a likelihood of 
less than 1\% of being consistent with this line (displayed in red)
were rejected in an iterative manner to produce this fit.}
\vspace{10pt}
\label{halofit2}
\end{figure}

In Figure~\ref{halofit2} we show the result of fitting a model in which the velocity
dispersion is allowed to vary linearly with projected radius $R$.  The
systemic velocity was again fixed at $-300 {\rm km/s}$ for stability due to
a complex windowing function. 
We proceeded by first calculating the velocity
dispersions in individual fields, given the adopted window function, and assuming that the dispersion is constant 
and Gaussian at the position of the field. The resulting maximum-likelihood calculated values
are shown in the upper panel of Figure~\ref{halofit2}.
A single model with a linear dependence of velocity dispersion with radius
was then fit to the full dataset. 
The measurements for individual fields are {\it not} used in
the analysis, presented only to clarify what the maximum-likelihood algorithm is
latching onto.
Proceeding in an iterative manner, we rejected those fields where the
likelihood of the velocity dispersion conforming to the linear fit was less 
than 1\%, and re-computed the fit.
The likelihood contours of this model are shown in the lower panel of Figure~\ref{halofit2}; 
the most likely model has
$$ \sigma_v(R) = 152 - 0.90 {{R}\over{1\kpc}} \kms \kpc^{-1}.$$
At a distance of 10\,kpc, the inferred velocity dispersion of the halo is 
therefore $\sim143 \kms$, while at a distance of 60\,kpc 
the inferred velocity dispersion has changed to $\sim98 \kms$.  
These results stand in contrast to the essentially isothermal halo measured
for the inner Galaxy \citep{chiba00}; for M31, an isothermal model can be rejected 
at the $\sim 3\sigma$ level.

The individual field-by-field fits reveal that the
overall fit when we allow for a gradient in the dispersion is 
obviously dominated by the stars inside $40\kpc$.
While stars likely belonging to the Giant Stream have been removed from the halo
sample, several {\it stream} fields 
stand out very clearly as not belonging to the trend by having a very low velocity dispersion, 
as do fields in the {\it NE shelf} region (described in \citealt{ferguson02} which 
also shows stars with giant stream kinematic structure). The fact that these fields still stand
out from an otherwise well-behaved trend suggests that our windowing of the giant 
stream is not extensive enough. However, at the present time 
we lack a full understanding of the distribution of the stream debris over the face of M31,
forcing us to adopt this pragmatic approach.

The above two models are essentially just phenomenological descriptions of
the data, so in the following we attempt a more physically-motivated model.
Modern cosmology finds that the dark matter halos of galaxies have a
universal shape \citep[hereafter NFW]{navarro97}. Simulations show that
halo stars are the remnants of ancient accreted proto-galaxies, but ones
that were relatively massive. As such the stars deposited by these
accretions are biassed towards being closer to the centre of the halo, and
consequently the halo is less extensive than the dark matter halo
\citep{abadi05}.  Nevertheless, the halo stars provide a reasonable
first approximation to the dark matter, and we will therefore attempt to fit
the population with an NFW model. This comparison is further motivated
by the finding that the velocity dispersion of the halo sample decreases
with radius, which is expected from the NFW model.

\begin{figure}
\vspace{6pt}
\centerline{\psfig{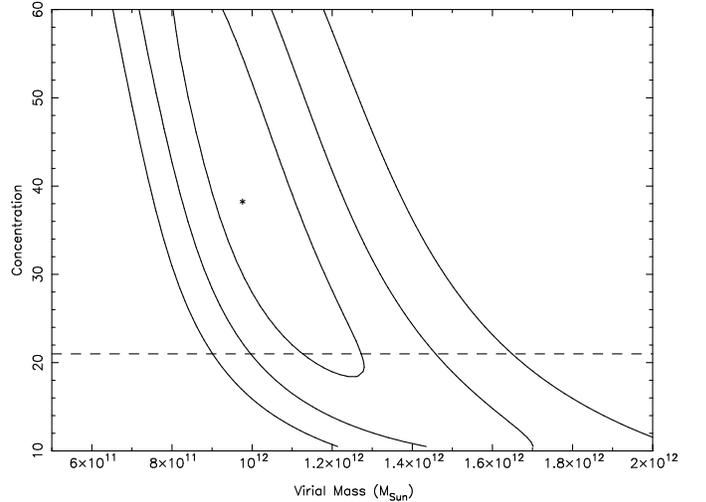}}
\vspace{6pt}
\caption{Likelihood contours (at $1\sigma$, $2\sigma$ and $3\sigma$
intervals) of an NFW model fit to our halo star sample, after rejection
of the low-dispersion fields identified from the analysis presented in Figure~6. Here the most
likely model has a virial mass of M$_V = 9.8 \times 10^{11} \msun$ and
concentration $c=38$. Notice however, that the constraint on the model
concentration provided by our analysis is not strong, and substantially
lower concentration values are allowed (lower than $c=18$ within the
$1\sigma$ boundary).  The dashed line shows the $c=21$ upper limit to the
concentration of an M31-like galaxy argued by \citet{klypin02}. Using this
additional constraint, the virial mass of the M31 halo must be greater than
M$_V = 9.0 \times 10^{11} \msun$ with 99\% confidence.}
\vspace{10pt}
\label{halofit3}
\end{figure}

The NFW model is fully described by concentration and virial mass
parameters. Though the concentration is on average dependent on the virial
mass \citep{bullock01}, there is a large scatter in this relation, and it is
therefore safer to consider the full parameter space.  We integrated the NFW
model numerically, using the relations described in \citet{lokas01}, to
obtain the projected line of sight velocity dispersion for each field in our
survey, under the assumption that the velocity dispersion is isotropic. Note
that the relative densities between fields were purposefully not specified,
due to the fact our survey does not provide a complete census of stars in
each field.  We further adopted $H_0 = 70 \kms$ and a virial overdensity
parameter of 200.  Figure~\ref{halofit3} shows the resulting likelihood
contours (after rejection of the low-dispersion fields marked red in the upper panel of 
Figure~\ref{halofit2}). 
Although the most likely model parameters are M$_V = 9.8 \times
10^{11} \msun$ and $c=38$, the likelihood contours are very extended in the
concentration direction. The main reason for this is that our
maximum-likelihood analysis method, as mentioned before, does not have
access to the information on the field-to-field density variations, whereas
the concentration by its very nature relates to the radial variation in
density. What density sensitivity we do have is due to the integration along
the line of sight in each field.  Cosmological simulations indicate that
concentration values of $c > 21$ are very unlikely for the dark halos of
galaxies such as M31 \citep{klypin02}; this limit is shown by a dashed line
in Figure~\ref{halofit3}. Imposing this additional constraint implies that the virial
mass of the M31 halo must exceed M$_V = 9.0 \times 10^{11} \msun$ with 99\%
confidence.  This value for the halo mass agrees remarkably well with the
value calculated from the kinematics of the giant stream around M31 \citep{ibata04},
where we found a mass interior to $125\kpc$ (the apocenter of the stream) of
$M(R < 125\kpc) = 7.5^{+2.5}_{-1.3} \times 10^{11} \msun$. In contrast, the
present constraint translates to $M(R < 125\kpc) > 7.2 \times 10^{11}
\msun$.

\section{Results: Metallicity properties of the stellar halo}

The photometrically dominant extended rotating component appeared to have a
relatively high metallicity, [Fe/H]=-0.9 $\sigma$=0.2 dex (I05), consistent with our
larger sample of disk stars reported here, [Fe/H]=-1.0 $\sigma$=0.3 dex
(where the error bars represent field to field dispersion in the 
measurements), explaining why so many disk-like stars have been found at 
large radius \citep{reitzel02,reitzel04}.  
While attempts to model a more metal-rich
halo have had some degree of success under specialized evolutionary
conditions \citep{font06}, our CDM understanding of how the halo of a
typical L$^{*}$ spiral galaxy forms suggest there must be a metal-poor relic
of the early halo.

But is there perhaps a metal-poor halo lurking underneath the relatively metal-{\it
rich} rotating component?  Our first goal is to assess the metallicity of
our velocity-windowed sample of stars, to see if it shows properties
distinct from the prominent extended rotating component which vastly
outnumbers the halo (on average by a factor $\sim6\times$) at radii
$<50$\,kpc \citep{irwin05}.

\begin{figure}
\vspace{6pt}
\centerline{\psfig{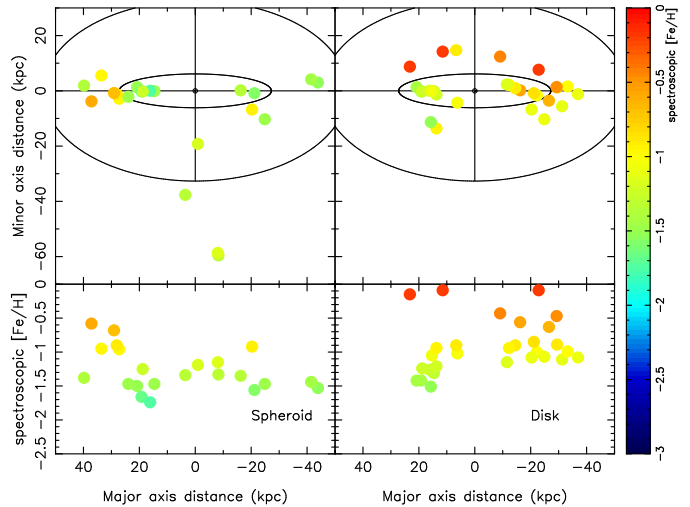}}
\vspace{6pt}
\caption{The metallicity measurements from stacked spectra are shown for the
halo sample (left panels) and disk sample (right panel).  The upper
panels show the spatial location of the fields, with the ellipses denoting
the same limits as in Figure~1. The smaller number of fields compared to
Figure~1 is due to the requirement that at least ten well-measured spectra
of the same galactic component be present in a field to provide a
metallicity measurement.
The five metal-rich halo fields in the North-East are not typical for the halo sample, and
are clearly the result of contamination from a [Fe/H]$\sim$-0.7 dex 
kinematically coherent structure 
at $\sim$-380\,km/s which appears in all these fields, 
likely to be associated with the giant southern stream \citep{chapman06}.
}  
\vspace{10pt}
\label{metalfields}
\end{figure}

\begin{figure*}[htb]
\vspace{6pt}
\centerline{\psfig{figure=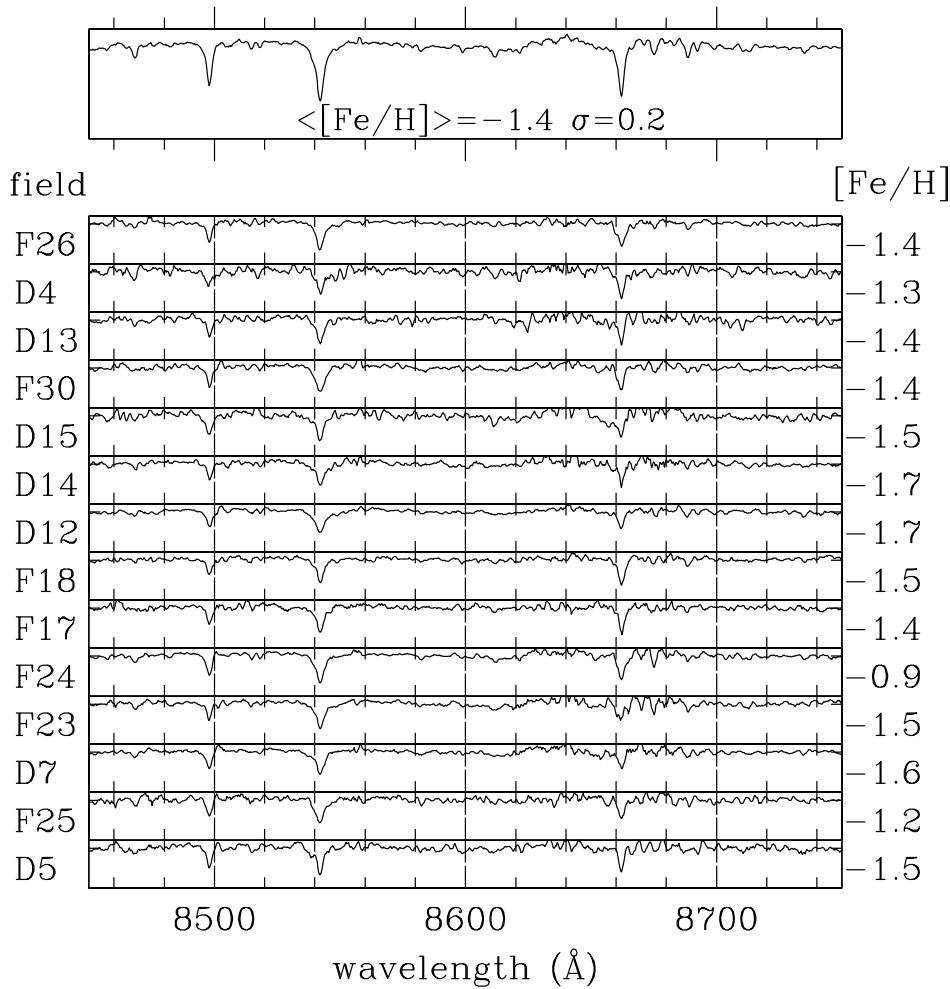,angle=0,width=5.5in}}
\caption{
Representative combined spectra (weighted by the S/N of individual spectra in the stack)
for kinematically selected halo stars in 14 of the 23 fields shown in Fig.~\ref{metalfields}, 
along with the total stacked halo spectrum at the top 
exhibiting an average [Fe/H]=-1.4 $\sigma$=0.2 dex, 
the error bar indicating the field to field RMS variation (whereas 
measurement errors in the stacked spectra are much smaller).
Field names and metallicities are indicated on the left and right respectively.
The spectra have been normalized to a continuum of one with the Ordinate
crossing at zero, and smoothed to the DEIMOS instrumental resolution.
The three Ca~II triplet lines at 8498.02\AA,
8542.09\AA, and 8662.14\AA\ are the most obvious features of these
spectra, although many other real features are present. }
\vspace{10pt}
\label{metalstack}
\end{figure*}

The DEIMOS spectra allow a measurement of the metallicity of the targeted
stars from the equivalent widths (EW) of the Ca~II triplet absorption lines,
free from the model-dependency of photometric analysis.
While the noise in individual spectra is typically too high to yield a
useful measurement of the Ca~II equivalent widths
(whereby individual spectra can have vastly different systematic errors 
due to the inopportune location of sky lines for certain
heliocentric velocities), we estimate the average
metallicity by stacking the RGB star spectra in individual fields. We shift
each spectrum to zero velocity before stacking, normalizing the continuum to one using the
inter-CaT continuum regions to fit a spline function, and finally weighting individual spectra
by the continuum S/N (Figures~\ref{metalfields} \& \ref{metalstack}).  Metallicities
were measured in an identical way to I05, using the correlation of the CaII
equivalent width with [Fe/H] as calibrated by \citet{rutledge97}.

\begin{figure}
\vspace{6pt}
\centerline{\psfig{figure=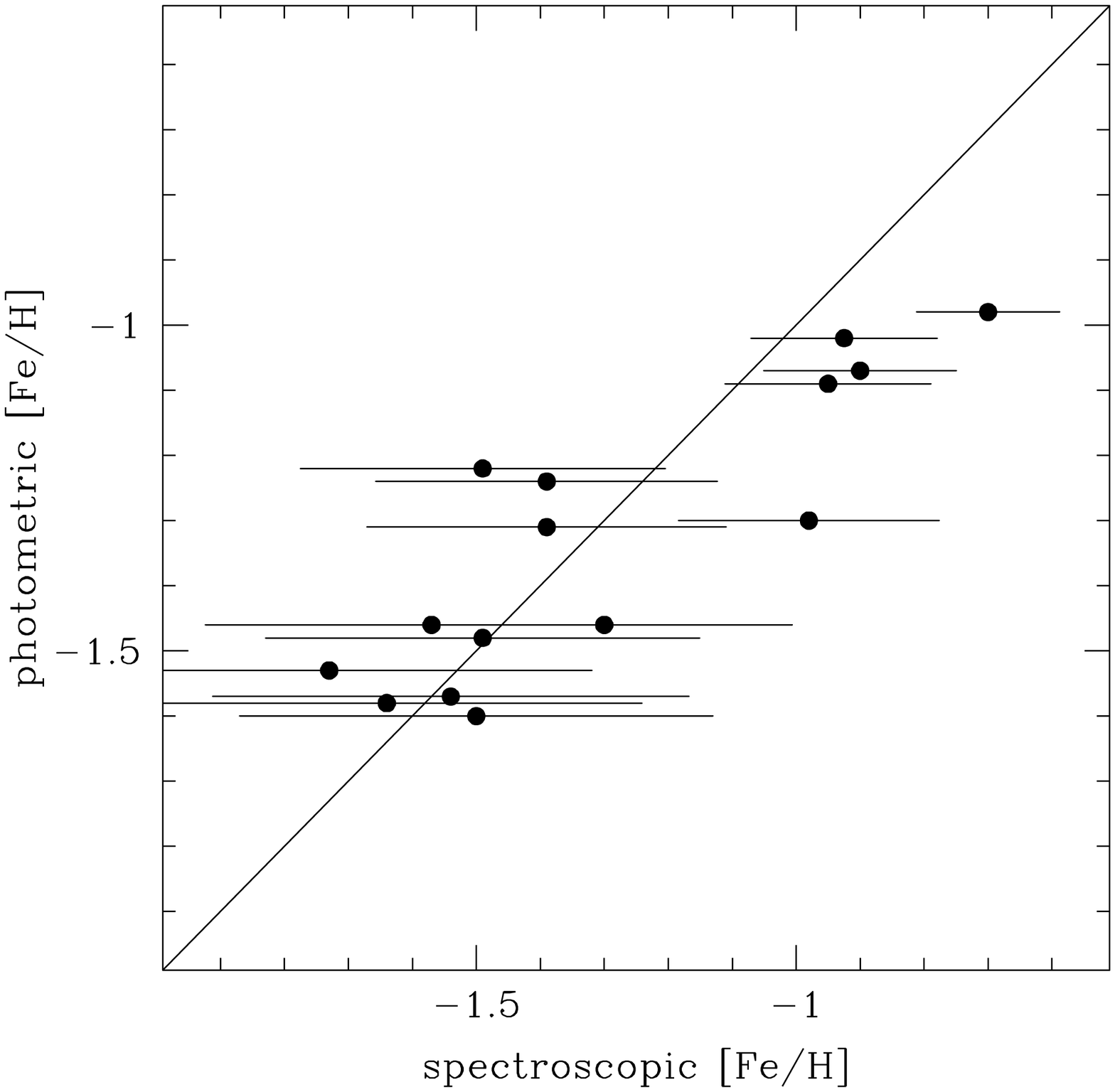,angle=0,width=3.5in}}
\vspace{6pt}
\caption{
Comparison of the spectroscopic and photometric [Fe/H] derived for RGBs in the
halo sample. Points are shown for all fields with at least 20 stars.
The line denotes equality. Note that the photometric [Fe/H] is highly
age dependent; the good agreement with the spectroscopic [Fe/H] suggests the 
adopted $\sim$13\,Gyr ages adopted from the globular cluster fiducials 
is reasonable for the halo stars.
The metal-rich points correspond to fields in the N-E region of the 
extended disk where the
giant southern stream (Ibata et al.\ 2001) passes through at velocities within the 
halo window. 
}  
\vspace{10pt}
\label{specphot}
\end{figure}

For comparison, we derived photometric [Fe/H] values for the halo stars by
interpolating between globular cluster fiducial sequences described in I05.
The photometric properties of these metal-poor halo RGBs are entirely
consistent with the spectroscopic properties, as shown in Figure~\ref{specphot}
(we defer the model/age-dependent
analyses of the stellar photometry to a followup paper).
The largest variations occur in the N-E fields where the metal-rich giant stream structure 
is present at the halo window velocities (v$_{hel}<-300$km/s), which appear to be more
metal-rich in the spectroscopic measurements. This may be indicative of the 
different ages/properties of the giant stream stars from the stellar halo.

The average CaT triplet metallicity in stacked spectra with at least 10 high quality halo
stars, is [Fe/H]=-1.4\ $\sigma=0.2$ (where the error quoted is simply the
dispersion in the individual field measurements), much poorer than we find
for the extended rotating component defined in the same manner
[Fe/H]=-1.0\ $\sigma=0.3$.  
There are five halo fields in the North-East which appear significantly more metal-rich than 
average and are not typical for the halo sample.
These five fields all possess a kinematically coherent sub-structure at 
v$_{hel}\sim-380$\,km/s and with [Fe/H]$\sim$-0.7 which contaminates
the halo-selected sample to various degrees. Including these sub-structure stars in the halo
sample only marginally changes the result ([Fe/H]=-1.36\ $\sigma=0.32$ with,
versus [Fe/H]=-1.40\ $\sigma=0.21$ without).
This sub-structure is likely to be associated with the giant southern stream 
\citep{ibata01b,ibata04}.

Aside from these {\it giant stream} fields in the North-East, 
the metallicity measurements in each halo field are
very similar, with no apparent radial gradient (Figure~\ref{metalfields}). 
The three most outlying metallicities found are for fields D14 and D12 (both with 
halo [Fe/H]$\sim$-1.7) and for F24 (with halo [Fe/H]$=$-0.9), none of which can
obviously be attributed to selection effects or contamination.
Fields D14 and D12 are in the N-E hemisphere of M31, where as we've noted the halo sample
cut with v$_{hel}<-300$km/s results in $<1$\% Galactic contamination.
While a small fraction of Galactic contamination is expected for 
F24 in the S-E hemisphere, this
would in fact make the spectroscopic [Fe/H] appear more poor, not rich.
These fields may therefore host accreted substructures which are not as well phase mixed
as the halo in general.  
Our findings therefore suggest
that there is a metal-poor, non-rotating halo component in M31, with
comparable metallicity to the Milky Way's non-rotating stellar halo
\citep{chiba00, chiba01}.

The accuracy of our M31 halo metallicity measurement is limited by the
continuum fitting in these relatively low signal-to-noise spectra, with
additional uncertainty from systematic effects such as residual sky.  The
large distance of M31 implies a small uncertainty for the horizontal branch magnitude, 
even if the stars are drawn from relative extrema in the galaxy.
Furthermore, the RMS dispersion in $V$-band magnitudes of the total sample
of halo stars is very small ($<$0.5 mag).  These effects diminish the impact
on the derived metallicity
of an uncertainty in distance in the RGB population we are observing in our
sample.  A formal error for each fit is obtained from the signal-to-noise of
the CaT lines, with an extra error estimated for the uncertainty in sky
subtraction.  The typical uncertainty is $\sim0.3$~dex.

A concern is that the average CaT metallicity in a given field may be
dominated by a few outlying spectra, although this should only be a factor
for increasing the metallicity (increasing the EW).  This potential
systematic error can be tested by dividing up the fields into halves and
comparing metallicity measurements. Generally subsamples have metallicities
consistent to within 0.2~dex.  However, a danger with this approach is that
for a small enough sample, variations will be present due to the random
sampling of the underlying metallicity distribution.

Our sample of RGB stars isolated to have halo kinematics
(and importantly windowing out the extended rotating component) have allowed
a census of the halo along both Major and minor axis fields, and underlying
the range of photometric substructures observed. 
The uniformly low (${\rm [Fe/H]} = -1.4$\ $\sigma$=0.2) 
metallicity we measure across the fields reflects 
a stellar halo component which has a distinctly different origin from the
extended rotating component.
Our discovery is complemented by the 
independent discovery of a halo component with 
metal poor ([Fe/H] = $-1.5$) stars lying 
at large radius (60--150~kpc) \citep{kalirai06, gilbert06}. 
Their outer halo component and bridging halo/bulge component at smaller radii 
are similarly metal-poor to the halo we describe here. 
However, the 47 pure halo stars at $R>60$\,kpc 
in \citet{kalirai06} and \citet{gilbert06} lack 
sufficient statistics to ascertain whether they are consistent with an outer extension of
the kinematically characterized, pressure-supported,
non-rotating component we describe in this paper.

The larger
field to field halo metallicity dispersion ($\sigma \sim 0.3$) found by \citet{kalirai06} and
\citet{gilbert06} is suggestive of 
an increasingly inhomogeneous halo component with radius, 
and is consistent with the dramatic structural 
variations from wide-field photometric surveys of 
M31's outer halo by Ibata et al.\ (2006) and \citet{martin06}.
These findings are also consistent with the much less phase-mixed outer halo in the
simulations of \citet{bullock05}, suggesting that the outer halo may truly represent
a component with a much longer assembly and relaxation time than the inner halo.

\section{Discussion}

While our study has uncovered conclusive evidence for a pressure
supported metal-poor halo in M31 within the inner 70\,kpc
(complemented by the co-discovery of a metal-poor halo in 
Kalirai et al.\ 2006 and Gilbert et al.\ 2006), evidence for
metal-poor stars in Andromeda's halo reaches back to \citet{holland96}, who
used deep imaging in the outer halo to probe the halo CMD.  Even earlier
results from RR~Lyrae studies suggested that old, metal-poor stars might be
present in the halo \citep{pritchet88}.  Metal-poor RGB stars were first
identified spectroscopically in the halo using Keck \citep{reitzel02,
reitzel04}, with speculations 
that differences appeared to exist between subsets of
the RGBs in their sample, finding a large range in metallicities, 
and suggesting that
disk-like stars were found at radial distances as large as 34\,kpc 
(de-projected).
However, with only two isolated pointings and small numbers of stars (23 and 29
respectively), these studies had no means to characterize the outer structural
components in M31.

While our Keck spectroscopic data do not suffer from age/metallicity
degeneracy as do the earliest photometric studies of the halo, the most
important ingredient for our discovery is the detailed knowledge of the
extended rotating component which allows for the separation of populations
by their kinematics.  Thus when stars are selected in a similar way to the
halo star selection in Galactic studies (i.e.\ by extreme kinematics --
\citealt{chiba00}), a similar [Fe/H]$\sim$-1.4 metallicity is found.

Since M31 and the Milky Way reveal so many differences in their evolutionary
histories and environments (e.g., Bulge size, thin disk scale, number and types 
of globular clusters, scale sizes and types of satellite dwarf galaxies -- 
e.g., \citealt{walterbos88, barmby00, huxor05, mcconnachie06a, mcconnachie06b}), it is of 
considerable interest that they seem to have developed similar metal-poor 
stellar components in their outskirts, likely near their formation epoch.  If models of halo formation are correct \citep{bullock05, renda05, font06}, this 
similarity in metallicity and dispersion, along with the similar dark matter masses \citep{evans00, ibata04}, suggest that both galaxies must have originated 
in comparable overdensity contrasts and perhaps attracted similar numbers and
masses of merger accretions early on. 

\begin{figure}
\vspace{6pt}
\centerline{\psfig{figure=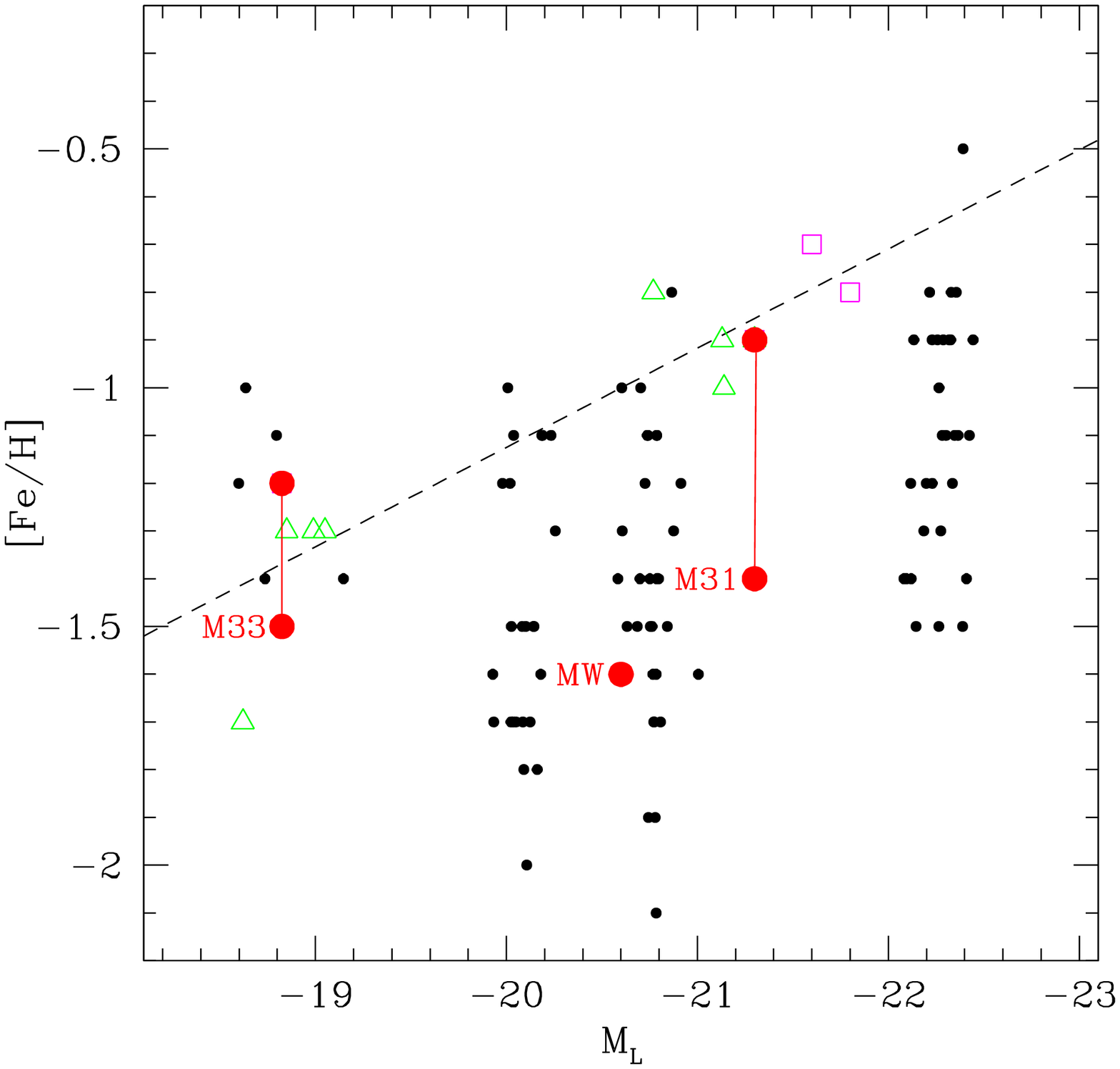,angle=0,width=3.5in}}
\vspace{6pt}
\caption{ The stallar halo metallicity - total galactic $V-$band luminosity relation.
Our kinematically selected metallicity measurements for M31 (this work) and
for M33 \citep{mcconnachie06c} are shown as revised points connected by lines
to literature values. 
Open triangles represent the HST data on other spiral galaxies from 
\citet{mouhcine05}; open squares represent
additional literature points extracted from \citet{renda05}.
Filled squares correspond to the peak of the metallicity distribution function (MDF)
in halo simulations of 
\citet{renda05}, where 68\% of stellar particle in the MDF are enclosed
within $\pm0.7$dex. The
ensembles at different luminosity ranges represent all the 
stellar particles in the simulation at a projected
distance $R>15$\,kpc, and corresponding to total masses (including dark matter)
of $1e11, 5e11, 1e12, 5e12 {\rm M}\odot$ respectively.
}

\label{renda}
\vspace{10pt}
\end{figure}

Our finding of a low halo metallicity in M31 contrasts, however,
with recent measurements of the stellar halo metallicity distribution
in a sample of more distant edge-on galaxies by \citet{mouhcine05}.
These authors derive photometric metallicity estimates from HST/WFPC2
observations and find a strong correlation between halo [Fe/H] and
galaxy luminosity (Figure~\ref{renda}).  
For a luminosity similar to M31 or the Milky Way,
their relation predicts a mean halo abundance of [Fe/H]$\sim-0.8$, 
significantly more metal-rich than that found here. It is important to point out
that the fields studied by \citet{mouhcine05} are assumed to
probe the stellar halo on the basis of their  projected
radii alone but this assumption has not been verified by either
structural decomposition, or kinematics.  We have shown that when
halo stars are selected kinematically -- arguably the best way in
which to do so -- then the metallicities inferred for the
Milky Way and M31 are both similarily low. When combined with the
recent discovery of metal-poor, [Fe/H]$\sim-1.5$, kinematically-selected
halo stars in the lower mass spiral galaxy M33 \citep{mcconnachie06c},
this suggests that the true underlying stellar halo metallicity in spiral galaxies may have
very little dependence on galaxy luminosity.

In Fig.~\ref{renda}, we place our
kinematically-selected halo metallicity results for M31 and M33
in the distribution of \citet{renda05}.
Their results suggest that galaxies with metal-rich stellar halos at $z=0$ 
have a longer formation history, 
whereas galaxies with a more metal-poor stellar halo at $z=0$ have a shorter
assembly, consistent with their halo [O/Fe] predictions.
By kinematically selecting only the non-rotating halo components of the Local Group
spirals,  it is clearly a shorter assembly time, primordial halo in the \citet{renda05}
simulations that would be isolated.
However, \citet{renda05} define the halo as all the 
stellar particles in the simulation at a projected
distance $R>15$\,kpc. Clearly a direct application of this definition to M31 would 
result in an average [Fe/H]$\sim$-1, being drawn primarily from the dominant
extended rotating component (I05). Note, however that the \citet{renda05} simulations
(Figure~\ref{renda}) suggest 
that even defining the total halo in this way, finding a halo metallicity as rich as 
[Fe/H]$\sim$-1 is quite rare for virial masses as large as M31 
($\sim10^{12}$M$_{\odot}$).
By contrast, in the \citet{font06} simulations
finding an early assembled stellar halo with a $10^{12}$M$_{\odot}$ virial mass and 
an [Fe/H]$\sim$-1.4 (like the MW and M31)
occurs in only $\sim$10\% of the accretion histories, suggesting that the MW and M31
halos are actually quite rare for L$^{*}$ spirals.


However, our discovery should not be interpreted as claiming that all
components of the $R<70$\,kpc stellar halo are metal-poor, but simply that a
non-rotating metal-poor component to M31 exists which is similar in [Fe/H]
to the halo in the Galaxy.  There may also be a rotating halo sub-component
which is more metal-rich which could effectively be windowed out of our
current sample.  Contamination from extended rotating component stars both
in velocity and metallicity makes a detailed comparison difficult.
\citet{ferguson02} and \citet{irwin05} noted that there is very clearly a
persistent population with red colors (suggesting stars which may be 
metal-rich or young) 
along the minor axis, and beyond the region
that I05 have characterized as the extended rotating component.  While
measuring the spectroscopic properties of this population would be
interesting, the statistics in our minor axis fields are currently
insufficient to show any kinematic differences or even allow robust
metallicity measurements for the red and blue RGBs. Future expansions of the
dataset may allow the characterization of this population.

Finally, it is of interest to understand how our new metal-poor component
connects to the extended $R^{-2}$ halo component discovered photometrically along the
minor axis by \citet{irwin05}, and also discovered through spectroscopic 
confirmation of halo-like, metal-poor stars at very large radius ($60-150$kpc) 
\citep{kalirai06, gilbert06}.  Are
these all part of the same halo component, and is there a gradient in
kinematics and metallicity?  Again, it is difficult to assess, although the
fact that massive satellites can be more easily accreted on prograde orbits
(contaminating the disk, but leaving the spherical halo intact), suggests
that such an extension of the halo would be physically motivated.
Tentative evidence from this work and from \citet{kalirai06}
for a more heterogeneous outer
halo component distinct from the more phase mixed inner halo component
is broadly consistent with the expectations from theory \citep{bullock05}.

\section{Conclusions}
We have presented evidence for a metal-poor, [Fe/H]$=-1.4, \sigma=0.2$ dex, stellar
halo component in M31, by kinematically isolating 827 non-rotating stars from our survey
of $\sim$10,000 stars with radial velocities from Keck-II/DEIMOS.
The halo component is detectable at radii from 10\,kpc to 70\,kpc, and underlies a more
dominant extended rotating component. 

This metal-poor  halo component  has no detectable metallicity gradient, and is consistent
with an early and rapid formation period as suggested by the simulations 
of \citet{renda05}. The more metal-rich stellar halo components observed with HST
by \citet{mouhcine05} may be indicative of components accreted later over 
longer assembly times.

Our analysis has also allowed us to constrain the kinematic properties of
the M31 halo. The windowed halo sample shows no evidence
for rotation about any axis, though there is clearly a drop in the velocity 
dispersion of the population with distance away from the galaxy center.
Fitting an NFW model to these data and using the cosmologically-motivated
constraint that the model concentration $c < 21$, provides a lower limit to
the virial mass of the halo of ${\rm M}_V > 9.0 \times 10^{11} \msun$ (99\% confidence).

\acknowledgements

SCC acknowledges support from NASA.  GFL acknowledges support through ARC
DP0343508 and thanks the Australian Academy of Science for financial support
in visiting the Institute of Astronomy, Cambridge.
AM would like to thank J.\ Navarro and S.\ Ellison for financial support.
Data presented herein were obtained using
the W.\ M.\ Keck Observatory, which is operated as a scientific partnership
among Caltech, the University of California and NASA. The Observatory was
made possible by the generous financial support of the W.\ M.\ Keck
Foundation.

\begin{deluxetable}{lccccccccl}
\renewcommand\baselinestretch{1.0}
\tablewidth{0pt}
\parskip=0.2cm
\tablenum{1}
\tablecaption{Log of Keck/DEIMOS spectroscopic observations in the Andromeda galaxy.\\
Positions are in degrees away from 00h42m44.3s +41d16m09s, J2000.0.}
\small
\tablehead{
number&field&name&Xki&Eta&no.stars$^a$&no.halo$^b$&no.disk$^c$&no.gal$^d$&obs.run\\&
}
\startdata
1&F1&w11&-1.398&-1.670&108&17&76&15&sept.2002\\
2&F2&50Disk&-0.968&-1.682&266&26&184&56&sept.2004\\
3&F3&w72&-1.463&-0.989&95&6&72&17&sept.2002\\
4&F4&w80&-1.120&0.020&79&6&49&24&sept.2002\\
5&F5&57Halo&1.087&-0.919&113&6&24&83&sept.2004\\
6&F6&10Disk&-0.443&0.462&226&4&146&76&sept.2004\\
7&F7&59Halo&2.364&-1.488&96&3&13&80&sept.2004\\
8&F8&54Blob&-0.532&1.061&206&8&113&85&sept.2004\\
9&F9&w91&-0.297&1.288&96&10&60&26&sept.2002\\
10&F10&56Blob&1.398&0.187&191&12&97&82&sept.2004\\
11&F11&w42&1.380&0.395&333&85&49&199&sept.2002\\
12&F12&w95&0.530&1.713&122&20&38&64&sept.2002\\
13&F13&06Disk&1.392&1.453&176&57&13&106&sept.2003\\
14&F14&14Disk&1.041&1.807&76&9&2&65&sept.2004\\
15&F15&53Blob&1.943&1.997&156&34&16&106&sept.2004\\
16&F16&55Blob&1.716&2.881&151&2&14&135&sept.2004\\
17&D1&17Disk&0.518&0.173&175&12&112&51&sept.2004\\
18&D2&18Disk&0.693&0.739&203&22&12&169&sept.2004\\
19&D3&52Disk&0.715&0.896&207&19&17&171&sept.2004\\
20&D4&51Disk&0.879&1.085&190&19&22&149&sept.2004\\
21&D5&04Disk&0.862&1.272&204&22&19&163&sept.2003\\
22&S1&s01&1.993&-3.965&74&6&32&36&sept.2002\\
23&S2&s02&1.740&-3.524&74&3&44&27&sept.2002\\
24&S6&s06&0.712&-1.755&89&27&44&18&sept.2002\\
25&S8&s08&0.189&-0.945&187&69&83&35&sept.2002\\
26&S24&s24&1.370&-4.954&131&5&35&91&sept.2003\\
27&S26&s26&0.593&-2.809&147&5&64&78&sept.2003\\
28&S27&s27&1.133&-2.566&145&4&77&64&sept.2003\\
29&D6&101DiH&-0.903&-1.353&279&12&257&10&sept.2004\\
30&D7&102DiH&-0.903&-1.243&270&13&233&24&sept.2004\\
31&D8&104DiH&-0.765&-0.923&273&14&218&41&sept.2004\\
32&D9&105DiH&-0.716&-0.786&273&13&244&16&sept.2004\\
33&D10&106DiH&-0.693&-0.605&269&15&211&43&sept.2004\\
34&F22&107ExH&-1.397&-1.643&268&5&221&42&sept.2004\\
35&F23&108ExH&-0.536&-1.891&259&13&185&61&sept.2004\\
36&F24&109ExH&-0.531&-1.470&251&27&174&50&sept.2004\\
37&F17&110HaS&-2.177&-2.402&157&27&10&120&sept.2004\\
38&F18&111HaS&-2.108&-2.198&123&15&10&98&sept.2004\\
39&F25&123GlS&3.080&-3.161&104&3&20&81&oct.2005\\
40&F26&124GlS&3.048&-3.127&109&4&21&84&sept.2005\\
41&D12&131DiH&0.653&0.852&289&3&39&227&oct.2005\\
42&D14&134DiH&0.867&1.115&279&27&25&227&oct.2005\\
43&D15&135DiH&1.189&1.300&213&74&15&124&oct.2005\\
44&F27&137DiH&1.338&1.581&262&99&21&142&oct.2005\\
45&F28&138DiH&1.356&1.656&265&20&26&119&oct.2005\\
46&F21&148ExH&-1.080&-2.045&213&17&116&80&sept.2005\\
47&F29&150ExH&1.206&2.205&229&74&38&117&sept.2005\\
48&F30&151ExH&1.695&2.396&195&43&16&136&sept.2005\\
49&F31&152ExH&1.679&2.536&209&33&18&158&sept.2005\\
50&S28&153ExH&0.1916&-0.863&170&21&84&65&sept.2005\\
51&D13&165DiH&0.747&0.948&265&27&16&222&oct.2005\\
52&D11&166DiH&-0.6643&-0.553&240&20&132&88&oct.2005\\
53&F20&167HaH&-1.597&-1.838&209&16&128&65&oct.2005\\
54&F19&168HaH&-1.605&-2.180&170&19&35&116&oct.2005\\
\hline
total&{}&{}&{}&{}&         9861&        1207&        4032&        4622& {}\\
\enddata
\tablenote{Total number of stars with cross-correlation peaks greater than 0.05.}
\tablenote{halo stars: all stars lagging the peak of the disk velocity by
 v$_{\rm lag}>160$km/s, or when no disk is detected in outer halo fields, then from the
 average disk lag, $25$km/s.
 From this sample of 1207 RVs, the giant stream is windowed out, and only stars with velocity
 errors $<20$km/s are used, giving a final Halo sample of 827 stars.}
\tablenote{Disk stars (encompassing both thin and thick components, as described in the
text): stars rotating with our disk model with a velocity lag v$_{\rm lag}<160$km/s.}
\tablenote{Stars with radial velocities falling within the {\it Galactic window}, defined as
v$_{\rm hel}>-160$km/s. Note that this column is only for consistent book-keeping of the Halo sample
across fields, but many stars from this window are used for disk analyses (see e.g., I05, and
fig.~\ref{vels}). The actual Galactic fraction of the total sample is estimated at $\sim$25\%.}
\label{tab1}
\end{deluxetable}

\end{document}